\documentclass[conference]{IEEEtran}
\IEEEoverridecommandlockouts
\usepackage{cite}
\usepackage{amsmath,amssymb,amsfonts}
\usepackage{algorithmic}
\usepackage{graphicx}
\usepackage{textcomp}
\usepackage{enumitem}  
\usepackage[table]{xcolor} 
\usepackage{colortbl} 
\usepackage{caption} 
\usepackage[hidelinks]{hyperref}
\usepackage{multirow}
\usepackage{arydshln}  

\def\BibTeX{{\rm B\kern-.05em{\sc i\kern-.025em b}\kern-.08em
    T\kern-.1667em\lower.7ex\hbox{E}\kern-.125emX}}
\begin{document}


\title{Pseudo Strong Labels from Frame-Level Predictions for Weakly Supervised Sound Event Detection}

\author{
Yuliang Zhang, Defeng (David) Huang, Roberto Togneri \\
\IEEEauthorblockA{
\textit{School of Electrical, Electronic and Computer Engineering, The University of Western Australia} \\
yuliang.zhang@research.uwa.edu.au, \{david.huang, roberto.togneri\}@uwa.edu.au}
}

\maketitle

\begin{abstract}
Weakly Supervised Sound Event Detection (WSSED), which relies on audio tags without precise onset and offset times, has become prevalent due to the scarcity of strongly labeled data that includes exact temporal boundaries for events. This study introduces Frame-level Pseudo Strong Labeling (FPSL) to overcome the lack of temporal information in WSSED by generating pseudo strong labels from frame-level predictions. This enhances temporal localization during training and addresses the limitations of clip-wise weak supervision. We validate our approach across three benchmark datasets—DCASE2017 Task 4, DCASE2018 Task 4, and UrbanSED—and demonstrate significant improvements in key metrics such as the Polyphonic Sound Detection Scores (PSDS), event-based F1 scores, and intersection-based F1 scores. For example, Convolutional Recurrent Neural Networks (CRNNs) trained with FPSL outperform baseline models by 4.9\% in PSDS\textsubscript{1} on DCASE2017, 7.6\% on DCASE2018, and 1.8\% on UrbanSED, confirming the effectiveness of our method in enhancing model performance.

\end{abstract}

\begin{IEEEkeywords}
Sound Event Detection, Weakly Supervised Learning, Convolutional Recurrent Neural Networks, Pseudo Strong Labels
\end{IEEEkeywords}


\section{Introduction}
\label{sec:intro}

\begin{figure*}[t!]

    \begin{minipage}[b]{1.0\linewidth}
        \centering \centerline{\includegraphics[height=0.25\textheight,width=\textwidth]{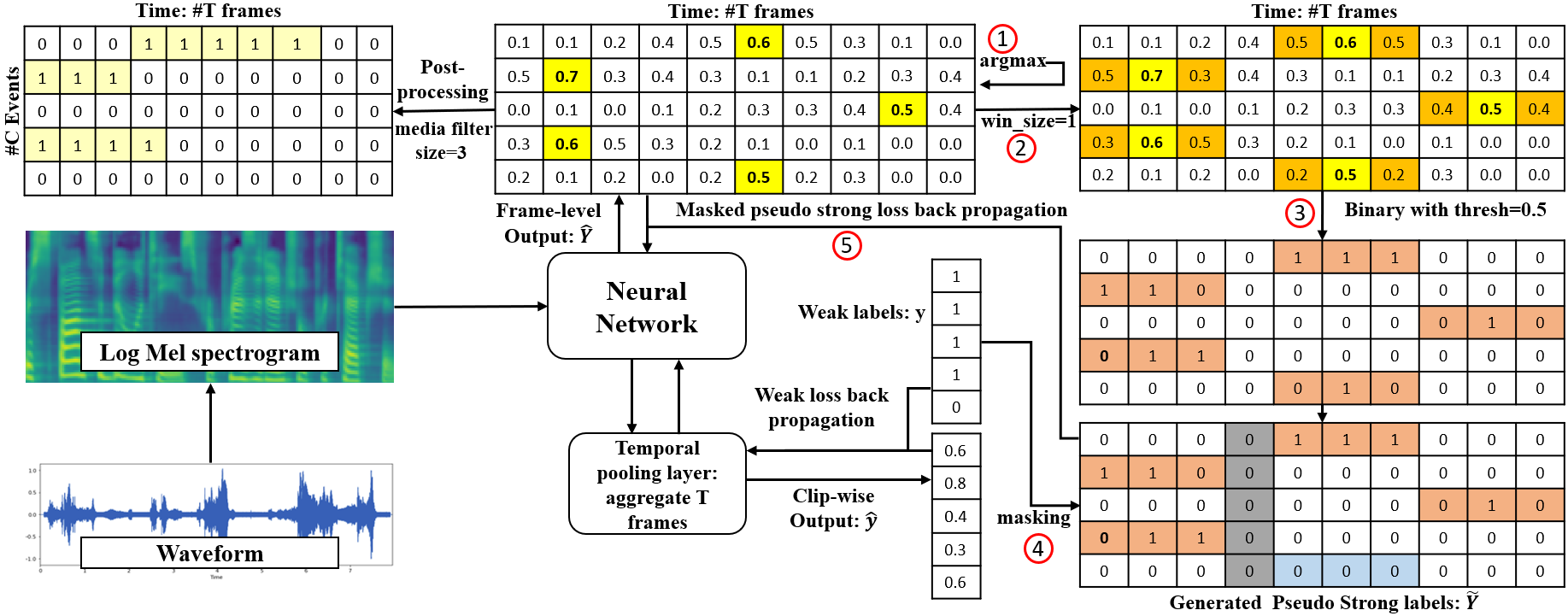}}
    \end{minipage}
    \captionsetup{skip=1pt} 
    \caption{Overall Architecture of the Frame-Level Pseudo Strong Label Training Procedure.}
    \label{fig:fig1_fpsl_framework}
\end{figure*}

Sound Event Detection (SED) is pivotal for applications such as surveillance, environmental monitoring, and smart home systems, capable of identifying sound events and their temporal boundaries in audio recordings \cite{c1, c2, c3, c4, c5, c6}. Typically, SED systems are trained using supervised learning, which relies on laboriously annotated data specifying the onset and offset of each sound event \cite{c7}. This process is time-consuming, leading to a scarcity of strongly labeled data.

Due to the limited availability of strongly labeled data, Weakly Supervised Sound Event Detection (WSSED) has received significant attention. These models operate under the constraint of having only audio tags for training, without precise temporal boundaries. The prevalent architectures in WSSED are based on Convolutional Recurrent Neural Networks (CRNNs) \cite{c8, c9, c10}. Notably, CDur \cite{c9} integrates duration modeling and temporal attention to enhance detection across variable event lengths. Similarly, FDY-CRNN \cite{c10} employs a frequency-adaptive convolution framework that dynamically adjusts filter parameters across frequency bands, thereby capturing frequency-specific patterns and boosting detection accuracy.

In addition to diverse modeling approaches, various temporal pooling strategies have been developed to aggregate frame-level predictions into single clip-wise probabilities. McFee et al. \cite{c11} introduced the softmax pooling, which combines the advantages of the mean and the max pooling, to enhance the model's sensitivity and specificity. The temporal max-pooling, particularly effective in detecting rare sound events, has also been utilized \cite{c12}. Furthermore, attention pooling methods \cite{c13, c14} are employed to focus the model more acutely on significant frames, segments, and frequency components, thereby improving the overall accuracy of the SED system.

A principal challenge in WSSED is the lack of detailed temporal information for sound events. Kim et al. \cite{c15} showed that point labels, marking a single time point per event within a recording, significantly enhance model performance compared to weakly labeled models. Inspired by this, our paper introduces the Frame-level Pseudo Strong Labeling (FPSL), a novel approach that generates pseudo strong labels from frame-level predictions during training. Unlike \cite{c15}, which relies on \textit{ground-truth} point labels, our method innovates by integrating temporal prediction directly into the WSSED training process. This one-stage mechanism, distinct from prior methods focusing primarily on new models or pooling techniques, can be easily implemented across most WSSED models to improve the precision of temporal detection.

Despite advances in WSSED model architectures and pooling methods, the lack of detailed temporal information in weak labels remains a significant performance barrier. The proposed FPSL mechanism can effectively addresses this gap by identifying likely time points for sound events during training, enhancing the model's ability to accurately delineate temporal boundaries. This method integrates both weakly supervised and pseudo strong supervised losses, ensuring the model focuses on audio tagging while capturing the temporal nuances of sound events, particularly for non-stationary, short-duration events. Our evaluations across three benchmark datasets—DCASE2017 Task 4, DCASE2018 Task 4, and UrbanSED—demonstrate that models trained with FPSL significantly outperform those without it. For instance, Convolutional Recurrent Neural Networks (CRNNs) trained with FPSL show improvements of 4.9\%, 7.6\%, and 1.8\% in PSDS\textsubscript{1} on the DCASE2017, DCASE2018, and UrbanSED datasets, respectively.

This paper is organized as follows: Section 2 describes our proposed approach, Section 3 details the experimental setup, Section 4 presents the results and analysis, and Section 5 offers the conclusion.

\section{Proposed Method}
\label{sec:method}

\subsection{Pseudo Strong Label Construction}

The FPSL framework is illustrated in Fig. \ref{fig:fig1_fpsl_framework}. Each frame-level output encapsulates $T$ time frames, with each frame related to $C$ distinct sound events. During each training epoch, we perform the following procedures to construct the FPSL:

\subsubsection{Locating Maximum Values}

For each sound event class $c$, identify the frame time $t^*_c$ with the maximum value in the frame-level output:

\begin{equation}
    t^*_c = \arg\max_{t \in \{1, \dots, T\}} Output(c, t)
\end{equation}
where $ Output(c, t)$ denotes the predicted probability of class \( c \) at time \( t \), shown in the top-middle table of Fig. \ref{fig:fig1_fpsl_framework}. The frames with maximum values are highlighted in yellow in the top-middle table of Fig. \ref{fig:fig1_fpsl_framework}

\subsubsection{Extending Frame Times }

The identified frame times $t^*_c$ are extended on both sides by a window size $win\_size$, as illustrated in orange in the top-right table of Fig. \ref{fig:fig1_fpsl_framework}. This extension is necessary because sound events typically span multiple frames, and the neighboring frames often have high probabilities when an actual sound event occurs:
\begin{equation}
    Extended(t^*_c) = \{ t^*_c - win\_size, \dots, t^*_c + win\_size \}
\end{equation}
Ensure the time indices stay within the range $1 \leq t \leq T$

\subsubsection{Applying Confidence Threshold}

Apply a binary confidence threshold $thresh$ to generate the FPSL as depicted in the middle-right table of Fig.\ref{fig:fig1_fpsl_framework}. For each extended frame time $t$ in $Extended(t^*_c)$:
\begin{equation}
    FPSL(c, t) = 
    \begin{cases} 
    1 & \text{if }  Output(c, t) \geq thresh \\
      &  for \text{ } t \in Extended(t^*_c) \\
    0 & \text{otherwise}
    \end{cases}
\end{equation}
\subsubsection{Masking with Weak Labels}

Use the weak labels $WeakLabel(c)$ as ground truth to mask the pseudo strong labels. If the weak label for a class $c$ is 0, all corresponding pseudo strong labels are masked as 0, illustrated by the blue cells in the bottom-right table of Fig. \ref{fig:fig1_fpsl_framework}.
\begin{equation}
    FinalFPSL(c, t) = FPSL(c, t) \times WeakLabel(c)
\end{equation}

\subsubsection{Generating Back Propagation Mask}
During the backpropagation stage, only relevant frames are utilized to train the models. The purpose is to reduce the generated noisy labels as in \cite{c23}. Specifically, only if frame at $t$ contains a maximum value (i.e., $t=t^*_c$ for some $c$) and there exists a $c$ that satisfies the condition $Output(c, t^*_c) \geq thresh$, then this frame and its adjacent frames within the $win\_size$ are included in the backpropagation process. For instance, the gray cells in the bottom-right table of Fig. \ref{fig:fig1_fpsl_framework} are excluded from the backpropagation process. The backpropagation mask $Y^M \in [0, 1]^{C \times T}$ is defined such that the values in the $t^{th}$ column are the same, given by: 


\begin{equation}
    Y^{M}(t) = 
    \begin{cases} 
    1 & \text{if } t \in Extended(t^*_c) \text{ and } \\ 
      &  \exists c \text{ such that } Output(c, t^*_c) \geq \text{thresh} \\
    0 & \text{otherwise}
    \end{cases}
\end{equation}

\subsection{Loss Function Combined with FPSL}

Binary cross-entropy loss is applied to both the weakly supervised component, represented by weak labels $y \in \{0, 1\}^{C}$, and the pseudo-strong supervised component, represented by FPSL labels $\tilde{Y} \in \{0, 1\}^{C \times T}$. Frame-level outputs are denoted by $\hat{Y} \in [0, 1]^{C \times T}$, and clip-wise outputs by $\hat{y} \in [0, 1]^{C}$. The loss function formulation is detailed in equation (\ref{eq: loss_func}).
\begin{equation}
    \label{eq: loss_func}
    Loss = L_{weak}(y, \hat{y}) + \alpha L_{FPSL}(\tilde{Y} \odot Y^{M}, \hat{Y} \odot Y^{M})
\end{equation}
where $\alpha$ is a hyper-parameter to determine the contribution of the pseudo-strong loss to the final loss, and $\odot$ indicates element-wise multiplication.

\section{Experimental Setup}
\label{sec:exp_setup}
\subsection{Datasets}

This paper evaluates the proposed method using three well-known datasets: DCASE2017 Task 4, DCASE2018 Task 4, and UrbanSED. The DCASE2017 Task 4 dataset was designed for smart car applications, which includes a large-scale collection of weakly supervised data, featuring 17 car-related sound events. It consists of 10-second audio clips, with 51,172 clips for training and 1,103 clips for evaluation; the validation subset was not used in this study. The DCASE2018 Task 4 dataset, focused on domestic environments, contains 1,578 weakly labeled 10-second clips for training and 1,168 strongly labeled clips for validation, used here for evaluation. URBAN-SED \cite{c18}, a synthetic dataset set in an urban context, comprises 10-second clips with 6,000 clips for training and 2,000 for evaluation, utilizing only the weakly labeled training and evaluation subsets in our experiments.

\subsection{Implementation Details \& Evaluation Metrics}

The input features for the SED model are log Mel spectrograms, extracted from 10-second audio clips sampled at 16 kHz, using a window size of 2048 samples and a hop length of 256 samples. We utilized two prominent CRNN-based architectures, CRNN \cite{c8} and FDY-CRNN \cite{c10}, to assess the effectiveness of the FPSL method. Attention pooling served as the temporal pooling layer for generating clip-wise predictions. Data augmentation techniques were applied to the DCASE2017 and DCASE2018 datasets. A uniform median filter size of 7 is used for post-processing across all datasets, although the top-left table of Fig. \ref{fig:fig1_fpsl_framework} depicts post-processing with a median filter size of 3. All other implementation specifics were adhered to as described in \cite{c10}. While we implemented the student and teacher model approach from \cite{c10}, no unlabeled datasets were used for training; the teacher model functioned solely as the exponential moving average of the student model.

The primary evaluation metrics used were the Polyphonic Sound Detection Scores (PSDS) \cite{c19}, including PSDS\textsubscript{1} and PSDS\textsubscript{2}, following the DCASE2021 Challenge Task 4 \cite{c20}. Additionally, event-based F1 scores (E-F1) and intersection-based F1 scores (IB-F1) with a threshold of 0.5 are provided \cite{c21, c22} for reference. The reported metrics represent the average results from three runs of the teacher model, each consisting of 200 training epochs, on the evaluation datasets.

\section{Results and Analysis}

In this section, we first validate the effectiveness and generalizability of the FPSL across three datasets using two advanced SED model architectures. We then conduct ablation studies on key hyper-parameters of FPSL, including $thresh$, $win\_size$, and $\alpha$. While performance differences exist across the three datasets, the overall trends remain consistent. To conserve space, we present representative ablation studies using the CRNN model on the DCASE2017 and DCASE2018 datasets, selecting the optimal model based on PSDS\textsubscript{1}. Finally, we discuss the efficacy of FPSL and provide guidelines for its application in WSSED.

\subsection{Effectiveness and Generalizability of FPSL}

\begin{table}[t!]
\centering
\captionsetup{skip=1pt} 
\caption{Average SED performance for three datasets and two models with $thresh=0.6$, $win\_size=1$, and $\alpha=0.3$.  A "+" indicates that FPSL has been applied.}
\resizebox{\linewidth}{!}{
\begin{tabular}{|c|c|c|c|c|c|c|}
\hline
\textbf{Dataset} & \textbf{Models} & \textbf{PSDS\textsubscript{1}} & \textbf{PSDS\textsubscript{2}} & \textbf{E-F1\textsubscript{mac}} & \textbf{E-F1\textsubscript{mic}} & \textbf{IB-F1} \\ \hline
\multirow{4}{*}{dcase2017} & crnn           & 0.124 & 0.281 & 0.101 & 0.103 & 0.284 \\ 
                           & crnn+      & \textbf{0.173} & \textbf{0.361} & \textbf{0.129} & \textbf{0.141} & \textbf{0.352} \\ \cline{2-7}
                           \cdashline{2-7}[1pt/1pt]
                           & fdy-crnn       & 0.104 & 0.255 & 0.083 & 0.095 & 0.286 \\ 
                           & fdy-crnn+  & \textbf{0.149} & \textbf{0.346} & \textbf{0.126} & \textbf{0.133} & \textbf{0.329} \\ \hline
\multirow{4}{*}{dcase2018} & crnn           & 0.222 & 0.490 & 0.187 & 0.262 & 0.459 \\ 
                           & crnn+      & \textbf{0.268} & \textbf{0.506} & \textbf{0.218} & \textbf{0.304} & \textbf{0.514} \\ \cline{2-7}
                           \cdashline{2-7}[1pt/1pt]
                           & fdy-crnn       & 0.261 & 0.506 & 0.179 & 0.203 & 0.456 \\ 
                           & fdy-crnn+  & \textbf{0.301} & \textbf{0.550} & \textbf{0.217} & \textbf{0.299} & \textbf{0.529} \\ \hline
\multirow{4}{*}{UrbanSED}  & crnn           & 0.074 & 0.073 & 0.199 & 0.196 & 0.475 \\ 
                           & crnn+      & \textbf{0.092} & \textbf{0.104} & \textbf{0.215} & \textbf{0.214} & \textbf{0.480} \\ \cline{2-7}
                           \cdashline{2-7}[1pt/1pt]
                           & fdy-crnn       & 0.075 & 0.024 & 0.210 & 0.208 & \textbf{0.486} \\ 
                           & fdy-crnn+  & \textbf{0.087} & \textbf{0.052} & \textbf{0.212} & \textbf{0.211} & 0.481 \\ \hline
\end{tabular}%
}
\label{table: general_res}
\end{table}

In this section, we set the parameters across all datasets and models with $thresh=0.6$, $win\_size=1$, and $\alpha=0.3$ for our experiments. While these settings may not be optimal for every dataset, they ensure the effectiveness of FPSL as a default configuration. Detailed results in Table \ref{table: general_res} show significant performance enhancements across all SED metrics with the inclusion of FPSL. Specifically, the PSDS\textsubscript{1} scores increased by 4.9\% on DCASE2017, 4.6\% on DCASE2018, and 1.8\% on UrbanSED with CRNN model. These improvements suggest potential for even greater gains with parameter tuning. Table \ref{table: general_res} consolidates these findings, underscoring FPSL’s effectiveness and broad applicability in SED tasks. The table highlights the benefits of FPSL, particularly in enhancing PSDS scores and other crucial SED evaluation metrics.

\subsection{Window Size in FPSL}

\begin{table}[t!]
\centering
\captionsetup{skip=1pt} 
\caption{Impact of window size on SED performance of crnn+ on DCASE2017 with $thresh=0.6$ and $\alpha=0.3$.}
\footnotesize 
    \begin{tabular}{|c|c|c|c|c|c|}
    \hline
    \textbf{Win\_Size} & \textbf{PSDS\textsubscript{1}} & \textbf{PSDS\textsubscript{2}} & \textbf{E-F1\textsubscript{mac}} & \textbf{E-F1\textsubscript{mic}} & \textbf{IB-F1} \\ \hline
    baseline  & 0.124 & 0.281 & 0.101 & 0.103 & 0.284 \\ \hline
    1  & \textbf{0.173} & 0.361 & 0.129 & 0.141 & 0.352 \\ \hline
    2  & 0.168 & 0.353 & 0.125 & 0.141 & 0.336 \\ \hline
    4  & 0.145 & 0.332 & 0.103 & 0.117 & 0.332 \\ \hline
    8  & 0.000 & 0.139 & 0.040 & 0.026 & 0.242 \\ \hline
    16 & 0.000 & 0.065 & 0.022 & 0.014 & 0.200 \\ \hline
    \end{tabular}
\label{table: win_size_impact_2017}
\end{table}

\begin{table}[t!]
\centering
\captionsetup{skip=1pt} 
\caption{Impact of window size on SED performance of crnn+ on DCASE2018 with $thresh=0.6$ and $\alpha=0.3$.}
\footnotesize 
    \begin{tabular}{|c|c|c|c|c|c|}
    \hline
    \textbf{Win\_Size} & \textbf{PSDS\textsubscript{1}} & \textbf{PSDS\textsubscript{2}} & \textbf{E-F1\textsubscript{mac}} & \textbf{E-F1\textsubscript{mic}} & \textbf{IB-F1} \\ \hline
    baseline  & 0.222 & 0.490 & 0.187 & 0.262 & 0.459 \\ \hline
    1  & 0.268 & 0.506 & 0.218 & 0.304 & 0.514 \\ \hline
    2  & 0.257 & 0.516 & 0.213 & 0.306 & 0.514 \\ \hline
    4  & \textbf{0.293} & 0.495 & 0.253 & 0.357 & 0.538 \\ \hline
    8  & 0.270 & 0.483 & 0.252 & 0.352 & 0.524 \\ \hline
    16 & 0.112 & 0.399 & 0.148 & 0.172 & 0.412 \\ \hline
    \end{tabular}
\label{table: win_size_impact_2018}
\end{table}

We fixed $thresh=0.6$ and $\alpha=0.3$, and varied the window size in powers of 2 to assess its impact. As shown in Table \ref{table: win_size_impact_2017}, system performance decreases as the window size increases.  Similarly, on the DCASE2018 dataset performance notably declines when the window size exceeds 4, as detailed in Table \ref{table: win_size_impact_2018}. These trends are consistent across other datasets and the FDY-CRNN model, although the specific points of change may vary slightly. Based on these observations, we recommend maintaining a smaller window size, ideally less than 5, to optimize performance.

\subsection{Confidence Threshold in FPSL}

\begin{table}[t!]
\centering
\captionsetup{skip=1pt} 
\caption{Impact of thresh on SED Performance of crnn+ on DCASE2017 with $win\_size=1$ and $\alpha=0.3$.}
\footnotesize 
\begin{tabular}{|c|c|c|c|c|c|}
\hline
\textbf{Thresh} & \textbf{PSDS\textsubscript{1}} & \textbf{PSDS\textsubscript{2}} & \textbf{E-F1\textsubscript{mac}} & \textbf{E-F1\textsubscript{mic}} & \textbf{IB-F1} \\ \hline
baseline  & 0.124 & 0.281 & 0.101 & 0.103 & 0.284 \\ \hline
0.3  & 0.165 & 0.346 & 0.123 & 0.144 & 0.332 \\ \hline
0.4  & 0.159 & 0.362 & 0.120 & 0.138 & 0.326 \\ \hline
0.5  & 0.163 & 0.359 & 0.123 & 0.137 & 0.353 \\ \hline
0.6  & \textbf{0.173} & 0.361 & 0.129 & 0.141 & 0.352 \\ \hline
0.7  & 0.161 & 0.363 & 0.130 & 0.142 & 0.344 \\ \hline
\end{tabular}
\label{table: thresh_impact_2017}
\end{table}

\begin{table}[t!]
\centering
\captionsetup{skip=1pt} 
\caption{Impact of thresh on SED Performance of crnn+ on DCASE2018 with $win\_size=1$ and $\alpha=0.3$.}
\footnotesize 
\begin{tabular}{|c|c|c|c|c|c|}
\hline
\textbf{Thresh} & \textbf{PSDS\textsubscript{1}} & \textbf{PSDS\textsubscript{2}} & \textbf{E-F1\textsubscript{mac}} & \textbf{E-F1\textsubscript{mic}} & \textbf{IB-F1} \\ \hline
baseline  & 0.222 & 0.490 & 0.187 & 0.262 & 0.459 \\ \hline
0.3  & \textbf{0.285} & 0.521 & 0.227 & 0.298 & 0.531 \\ \hline
0.4  & 0.279 & 0.529 & 0.223 & 0.298 & 0.524 \\ \hline
0.5  & 0.257 & 0.523 & 0.211 & 0.298 & 0.518 \\ \hline
0.6  & 0.268 & 0.506 & 0.218 & 0.304 & 0.514 \\ \hline
0.7  & 0.274 & 0.508 & 0.223 & 0.310 & 0.511 \\ \hline
\end{tabular}
\label{table: thresh_impact_2018}
\end{table}

We fixed $win\_size=1$ and $\alpha=0.3$, and varied the confidence threshold from 0.3 to 0.7 in increments of 0.1. The results, presented in Table \ref{table: thresh_impact_2017} and Table \ref{table: thresh_impact_2018}, demonstrate that in all settings, the model with FPSL significantly outperforms the baseline. Although the DCASE2018 dataset (see Table \ref{table: thresh_impact_2018}) achieved the best performance at $thresh=0.3$, we still recommend setting the confidence threshold between 0.5 and 0.7 to ensure improved performance of SED models in general.

\subsection{Contribution of FPSL in Loss Function}

\begin{table}[t!]
\centering
\captionsetup{skip=1pt} 
\caption{Impact of weight on SED performance of crnn+ on DCASE2017 with $thresh=0.6$ and $win\_size=1$.}
\footnotesize 
\begin{tabular}{|c|c|c|c|c|c|}
\hline
\textbf{Weight} & \textbf{PSDS\textsubscript{1}} & \textbf{PSDS\textsubscript{2}} & \textbf{E-F1\textsubscript{mac}} & \textbf{E-F1\textsubscript{mic}} & \textbf{IB-F1} \\ \hline
baseline  & 0.124 & 0.281 & 0.101 & 0.103 & 0.284 \\ \hline
0.1  & \textbf{0.178} & 0.359 & 0.144 & 0.147 & 0.344 \\ \hline
0.3  & 0.173 & 0.361 & 0.129 & 0.141 & 0.352 \\ \hline
0.5  & 0.170 & 0.360 & 0.119 & 0.135 & 0.341 \\ \hline
0.7  & 0.166 & 0.355 & 0.119 & 0.139 & 0.340 \\ \hline
0.9  & 0.177 & 0.364 & 0.119 & 0.142 & 0.342 \\ \hline
\end{tabular}
\label{table: weight_impact_2017}
\end{table}

\begin{table}[t!]
\centering
\captionsetup{skip=1pt} 
\caption{Impact of weight on SED performance of crnn+ on DCASE2018 with $thresh=0.6$ and $win\_size=1$.}
\footnotesize 
\begin{tabular}{|c|c|c|c|c|c|}
\hline
\textbf{Weight} & \textbf{PSDS\textsubscript{1}} & \textbf{PSDS\textsubscript{2}} & \textbf{E-F1\textsubscript{mac}} & \textbf{E-F1\textsubscript{mic}} & \textbf{IB-F1} \\ \hline
baseline  & 0.222 & 0.490 & 0.187 & 0.262 & 0.459 \\ \hline
0.1  & 0.248 & 0.505 & 0.211 & 0.300 & 0.500 \\ \hline
0.3  & 0.268 & 0.506 & 0.218 & 0.304 & 0.514 \\ \hline
0.5  & 0.278 & 0.527 & 0.216 & 0.306 & 0.528 \\ \hline
0.7  & 0.\textbf{280} & 0.522 & 0.228 & 0.320 & 0.530 \\ \hline
0.9  & 0.261 & 0.515 & 0.218 & 0.321 & 0.528 \\ \hline
\end{tabular}
\label{table: weight_impact_2018}
\end{table}

We fixed $thresh=0.6$ and $win\_size=1$, and varied $\alpha$ from 0.1 to 0.9 in increments of 0.2. Results presented in Table \ref{table: weight_impact_2017} and Table \ref{table: weight_impact_2018} demonstrate that models enhanced with FPSL consistently outperform the baseline across all settings of $\alpha$. Notably, setting $\alpha=0.1$ yielded the best performance in PSDS\textsubscript{1} and event-based F1 scores in DCASE2017, while setting  $\alpha=0.7$ results in the best performance in PSDS\textsubscript{1} and event-based F1 scores in DCASE2018. Further analysis with $thresh=0.5$ and $win\_size=4$ confirmed that a smaller $\alpha$ effectively maintains FPSL's efficacy. We recommend an $\alpha$ range of [0.1, 0.7] for general application.

\subsection{Analysis}

\begin{figure}[t!]
    \centering
    \captionsetup{skip=0pt} 
    \centerline{\includegraphics[width=8.5cm]{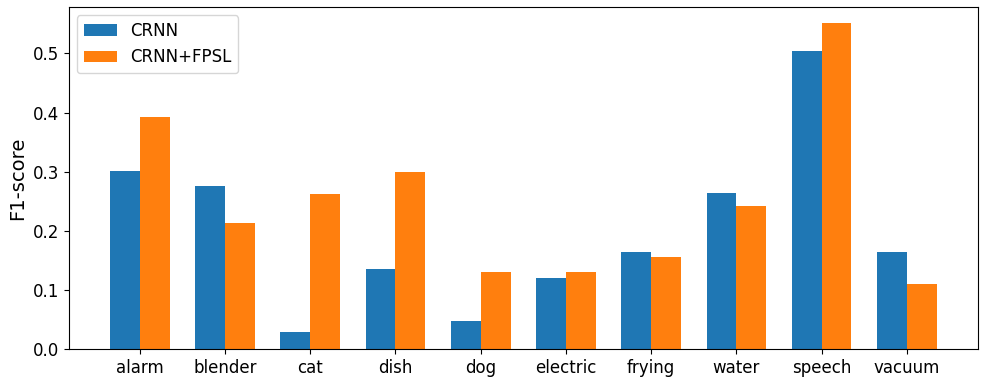}}
    \caption{Comparison of CRNN SED performance on the DCASE2018 dataset w/ and w/o FPSL in terms of class-wise event-based macro F1 score.}
    \label{fig: classwise_comparison}
\end{figure}

Intuitively, FPSL is expected to improve detection of short-duration sound events by enhancing the focus on key time points, but it may damage the performance of long, stationary sound events. Testing on the DCASE2018 dataset, which has varied event durations \cite{c9}, showed that with $thresh=0.5$, $win\_size=4$, and $\alpha=0.3$, FPSL achieved a PSDS\textsubscript{1} score of 0.298 using the CRNN model. Figure \ref{fig: classwise_comparison} confirms that FPSL enhances detection for short, non-stationary events (e.g., cats, dishes, dogs) but reduces performance on longer, stationary events (e.g., blenders, vacuum cleaners).

\begin{figure}[t!]
    \centering
    \captionsetup{skip=0pt} 
    \centerline{\includegraphics[width=8.5cm]{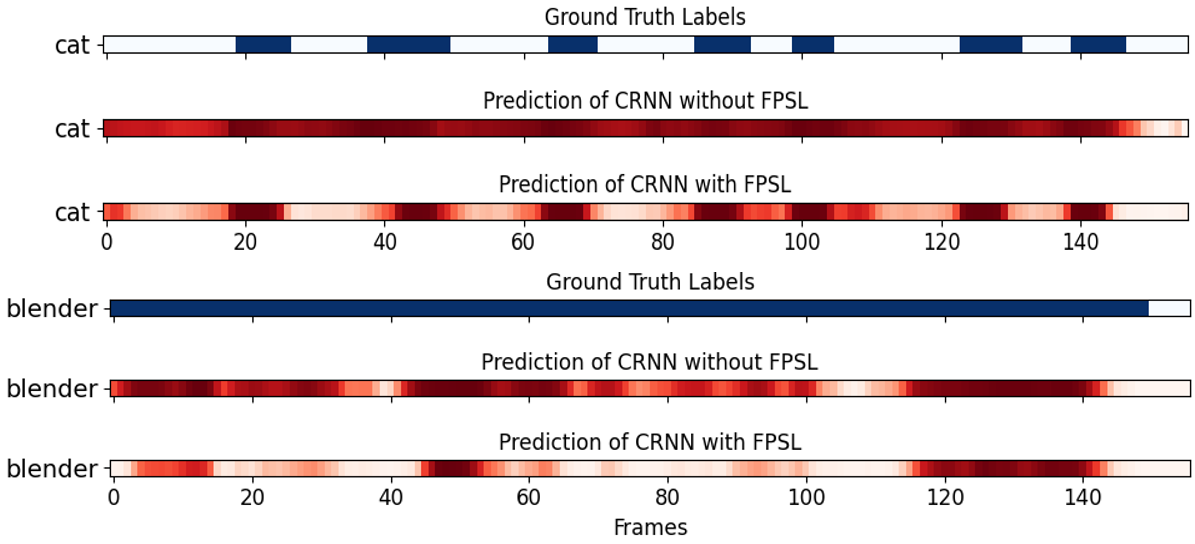}}
    \caption{Case study on "cat" and "blender" events in the DCASE2018 dataset: demonstrating the Pros and Cons of FPSL for non-stationary and stationary sound events}
    \label{fig: cat_and_blender}
    \vspace{-2ex} 
\end{figure}

Figure \ref{fig: cat_and_blender} further illustrates FPSL's effectiveness by comparing ground truth and predictions for 'cat' and 'blender' events with and without FPSL, demonstrating improved capture of critical temporal details but with a trade-off for stationary sound events.




\section{Conclusions}

Temporal accuracy is crucial for SED, but the lack of precise time boundaries in WSSED limits model performance. FPSL addresses this by directly integrating temporal information during training, significantly enhancing the detection of non-stationary, short-duration sound events. Experimental results across various datasets confirm FPSL's effectiveness, with optimal performance achieved using a confidence threshold of 0.5 to 0.7, a window size between 1 and 4, and an alpha value from 0.1 to 0.7. In the future, we will focus on improving FPSL to better track stationary sounds.

\vfill\pagebreak

\end{document}